\documentclass[prb,aps,reprint,superscriptaddress,showpacs]{revtex4-1}
\usepackage{amsmath,amssymb,graphicx}

\begin{document}
\title{Optically and electrically controlled polariton spin transistor}

\author{I. A. Shelykh}
 \affiliation{Science Institute, University of Iceland, Dunhagi-3, IS-107, Reykjavik, Iceland}
 \affiliation{International Institute of Physics, Av. Odilon Gomes de Lima 1722, 59078-400, Natal, Brazil}

\author{R. Johne}
\affiliation{Clermont Universit\'e, Université Blaise Pascal, LASMEA, BP 10448, 63000 Clermont-Ferrand, France; CNRS, UMR 6602, LASMEA, 63177 Aubi\`ere, France}
\affiliation{Eindhoven University of Technology, Applied Physics Institute, PO Box: 513, 5600MB Eindhoven, The Netherlands}

\author{D. D. Solnyshkov}
\affiliation{Clermont Universit\'e, Université Blaise Pascal, LASMEA, BP 10448, 63000 Clermont-Ferrand, France; CNRS, UMR 6602, LASMEA, 63177 Aubi\`ere, France}

\author{G. Malpuech}
\affiliation{Clermont Universit\'e, Université Blaise Pascal, LASMEA, BP 10448, 63000 Clermont-Ferrand, France; CNRS, UMR 6602, LASMEA, 63177 Aubi\`ere, France}

\date{\today}

\begin{abstract}
We propose two schemes of a novel spin-optronic device, optical
analog of Datta and Das spin transistor for the electrons. The role
of ferromagnetic contacts is played by one-dimensional polariton
channels with strong TE-TM splitting. A symmetric 2D
trap plays a role of the non-magnetic active region. The rotation
of the polarization of the pulse in this region can be achieved
either due to its interaction with a spatially confined polariton condensate
created resonantly by a circular polarized laser, either due to the splitting
between the two linear polarizations of the excitons, controlled
electrically by use of metallic gate.
\end{abstract}


\maketitle

A number of important phenomena in condensed matter physics are
connected with the spin of elementary excitations of different kinds:
electrons, holes, excitons, or exciton-polaritons. The possibility
to manipulate and control the spins of individual particles opens a
way to the implementation of spin degree of freedom in nano-
electronic devices of the new generation: spintronic devices (if
spin of electrons or holes is used \cite{Zutic2004}) and
spinoptronic devices (if spin of excitons or exciton polaritons is
used \cite{Shelykh2010}).

The first spintronic device was proposed in early 90ies in the
pioneer work of S. Datta and B. Das \cite{DattaDas}. It consists of
two ferromagnetic electrodes with collinear magnetizations,
separated by a non-magnetic semiconductor region with Rashba
Spin-Orbit Interaction (SOI). The Rashba Hamiltonian can be
interpreted in terms of an effective magnetic field in the plane of
a QW, perpendicular to the kinetic momentum of the carriers and
charachterized by  the Rashba parameter $\alpha$ which can be
efficiently tuned by varying the gate voltage $V_g$ applied to the
top gate metallic electrode covering the non-magnetic region
\cite{Nitta1997,Heida1998,Engels1997}. When the spin-polarized
carriers enter the non-magnetic region, this effective field
provokes the rotation of their spins. Depending on the rotation
angle $\Delta\phi=2m_{\text{eff}}\alpha L/\hbar^2$, the transmitted
current reveals periodic oscillations with minima and maxima
corresponding to angles $\Delta\phi=2\pi N$ and $\Delta\phi=\pi
(2N+1)$, respectively. Due to the problems of spin injection and
decoherence Datta and Das device remained for a long time the  a
purely theoretical concept, and it has been realized experimentally
only recently \cite{Koo2009}.

On the other hand, in the domain of mesoscopic optics it was
proposed to use exciton-polaritons for creation of
spinoptronic devices representing optical analogs of spintronic
devices. Exciton-polaritons (or cavity polaritons) are the
elementary excitations of semiconductor microcavities in the strong
coupling regime. Being composite particles consisting of bright
heavy-hole excitons and photons, polaritons have two allowed spin
projections on the structure growth axis ($\pm1$), corresponding to
the right and left circular polarisations of the counterpart
photons. Thus, from the formal point of view, the spin structure of
cavity polaritons is similar to the spin structure of electrons
(both are two-level systems), and their theoretical description can
be carried out along similar lines. It should be noted, however,
that the statistics of elementary excitations is different in two
kinds of systems: fermionic in the case of spintronics, bosonic in
the case of spin-optronics. Also, it appears that the account of
many-body interactions is of far greater importance for spinoptronic
devices with respect to the spintronic ones \cite{Shelykh2010}.

It was recently proposed that in the domain of spinoptronics the
problems of spin injection and decoherence can very probably be
resolved: the coherent length for polaritons is orders of magnitude
longer than that of electrons \cite{Langbein2007} and the spin of the injected polaritons
can be easily tuned by simple
change of polarization of the exciting laser.  The first
spinoptronic device, namely polarization controlled optical gate, was
recently realized experimentally \cite{Leyder2007}, and theoretical
schemes of several others were theoretically proposed. Among the
latter one should mention two proposals for optical analog of the
spin transistor: polariton Berry phase interferometer
\cite{Shelykh2009} and polariton Datta and Das device \cite{Johne}.
Both of them require strong magnetic fields for their operation,
which complicates their experimental realization and technological
implementations. In the present letter we propose two new schemes of
the polariton analog of Datta and Das device, which do not require
any magnetic field and where the tuning of the transmitted
polariton "current" can be achieved by purely optical or electrical methods.

The geometries of the two schemes are shown at Fig. 1 (a). In both
of them, the role of ferromagnetic contacts is played by 1D
polariton channels (polariton waveguides), where due to the strong
TE-TM splitting which is inversly proportional to the waveguide
dimension \cite{Dasbach2005} and plays a role of magnetic field
$B_0$ in ferromagnetic electrodes there is a region of the energies,
where the mode with TM linear polarization can propagate. In
realistic channels the value of the splitting can be as high as 1-2
meV for a waveguide of $1 \mu$m width \cite{Kurther1998}.
Technologically, polariton waveguiding can be achieved by etching
\cite{Dasbach2005}, variation of the cavity thickness
\cite{Idrissi2006}, applying stress \cite{Balili2006}, or putting
metals on the surface of the cavity \cite{Lai2007,Kaliteevskii2009}.
The 1D channels are connected to a symmetric 2D trap (\emph{active
region}), where the TE-TM splitting is absent. This trap plays a
role of the nonmagnetic region. An effective magnetic field acting
on the pseudospin of the pulse of the linear polarized polaritons in
the active region can be created by two alternative ways.

\begin{figure}
\begin{center}
\includegraphics[width=1.0\linewidth]{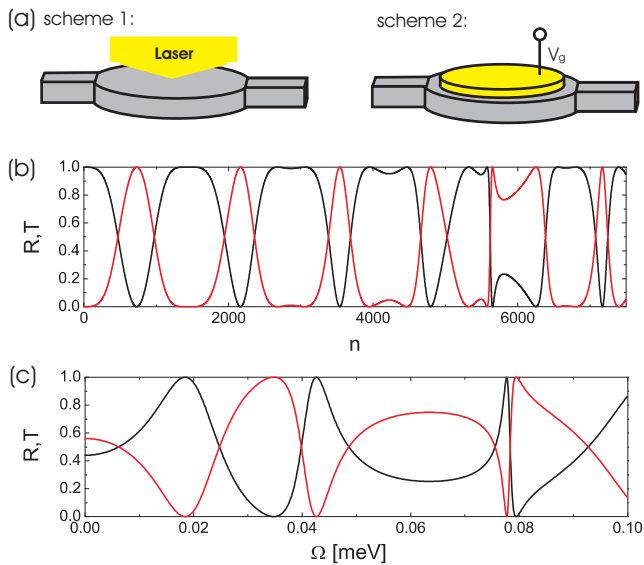}
\caption {\label{fig1} (a) Illustration of the optically (scheme 1)
and electrically (scheme 2) switched spin transistor. (b)
\textit{Scheme 1}: Transmission (black) and reflection coefficient of
the optically switched transistor in dependence on the number of polaritons $n_{cond}$ in the condensate with a wavevector of
$k=2\times10^6 m^{-1}$ of the incoming pulse. (c) \textit{Scheme 2}:
Transmission (black) and reflection coefficient of the electrically
manipulated transistor in dependence on the anisotropic polarization
splitting $\Omega \propto V_g$. The wavevector of the incoming pulse
$k=1.2\times10^6 m^{-1}$. The other parameters used in the
calculation are: $U_0=-1$ meV, $L=25$ $\mu$m and $\Delta_{LT}=10$
meV.}
\end{center}
\end{figure}

\emph{Scheme 1}. For optically controlled transistor, one creates a
circular polarized polariton condensate in the region of the trap by
a resonant cw pump. After entering this active region, the
polaritons from the pulse start interacting with polaritons of the
condensate. These interactions are strongly spin-anisotropic: the
interaction of polaritons in the triplet configuration (parallel
spin projections on the structure growth axis) is much stronger than
that of polaritons in the singlet configuration (antiparallel spin
projections) \cite{Renucci2005}. This leads to a mixing of linearly
polarised polariton states, manifesting itself in the appearance of
a z-directed effective magnetic field $B_{eff}$ acting on polariton
pseudospin. In the geometry we consider, this will lead to the
rotation of the pseudospin of the pulse along the z-axis. The
rotation angle can be estimated as
\begin{equation}
\Delta\phi=\frac{\Omega Lm_{eff}}{\hbar^2k} \label{RotationAngle}
\end{equation}
where L is the length of the active region, $m_{eff}$ is the
effective mass of the polaritons, $k$ is their in-plane wavenumber,
and $\Omega=(\alpha_1-\alpha_2)n_{cond}$ is a pump-induced splitting
in circular polarizations in the active region with
$\alpha_1,\alpha_2$ being the constants characterizing the
interaction of the polaritons in the singlet and triplet
configurations \cite{Glazov2009}, $n_{cond}$ being the 2D density of
circular polarized polariton condensate in a trap. In complete
analogy with the electronic Datta and Das transistor, due to the
strong TE-TM splitting in the leads one expects periodic dependence
of the polariton current on $\Delta\phi$. As $\Delta\phi$ depends on
the condensate density $n_{cond}$, the transmitted polariton current
can be modulated by tuning the intensity of the circular polarized
laser pump.

\emph{Scheme 2.} The rotation of the polarization of the pulse in
the active region can be also achieved by purely electrostatic
method via the application of a metallic gate electrode.
Variation of the gate voltage $V_g$ results in the changing of the
asymmetry of the QWs in the direction of the structure growth axis
$z$, which leads to the splitting in the energies of the excitonic
(and polaritonic) states with linear polarizations parallel to main
crystalline axes \cite{Aleiner1992,Malpuech2006} and appearance of
the corresponding effective in-plane magnetic field. If there is an
angle of $45^\circ$ between crystalline axes and polariton waveguides,
the effective magnetic fields in the waveguides $B_0$ and in the
active region $B_{eff}$ are perpendicular, and thus the pseudospin
of the linear polarized pulse entering the active region will
undergo the rotation by the angle given again by
Eq.\ref{RotationAngle}, the only difference being that now $\Omega$
is the splitting in the active region induced by the top gate, dependent on
$V_g$. The tuning of the gate voltage will thus result in the
periodic modulation of the outgoing polariton current.

For quantitative description of the polariton transport in the
system considered, we first use a simplified model, treating the
polariton waveguides and the active region as quasi-1D. The transmission and reflection amplitudes in this case can be found from the solution of the one- dimensional scattering problem with the following Hamiltonian written in a basis of circular polarizations

\begin{eqnarray}
\widehat{H}=\left\{ {\begin{array}{*{20}{c}}
   { - \frac{{{\hbar ^2}}}{{2m}}\frac{{{d^2}}}{{d{x^2}}} - \frac{{{\Delta _{LT}}}}{2}{\sigma _x}}  \\
   { - \frac{{{\hbar ^2}}}{{2m}}\frac{{{d^2}}}{{d{x^2}}} -U_0- \frac{\Omega }{2}{\sigma _{y,z}}}  \\
   { - \frac{{{\hbar ^2}}}{{2m}}\frac{{{d^2}}}{{d{x^2}}} - \frac{{{\Delta _{LT}}}}{2}{\sigma _z}}  \\
\end{array}} \right.
\label{Hamiltonian}
\end{eqnarray}
where we divide the system into three regions: (1) $x<0$ (ingoing waveguide),
(2) $0<x<L$ (active region) and (3) $x>L$ (outgoing waveguide). The axis x was chosen to be parallel to the direction of the leads, in the intermediate region one should use the term $\Omega/2\sigma _{z}$ for the scheme 1 and $\Omega/2\sigma _{y}$ for the scheme 2 with $\sigma_j$ denote Pauli matrices.  $m$- polariton effective mass, $\Delta_{LT}$- longitudinal- transverse splitting in the leads, $\Omega_{LT}$- effective magnetic field in the active region 2, $U_0$ the energy offset between polaritons in the waveguides and active region appearing from the confinement effects. The solutions of the stationary Schrodinger equation $\widehat{H}\Psi=E\Psi$ with a wavefunction $\Psi$ being a spinor and Hamiltonian \ref{Hamiltonian} in all three regions read:

\begin{eqnarray}
\label{wavefunction1}
\Psi_1=&(e^{ik x}+ r e^{ik x}) \frac{1}{\sqrt{2}}\begin{pmatrix} 1 \\ 1 \end{pmatrix} + A e^{\gamma x} \frac{1}{\sqrt{2}}\begin{pmatrix} -1 \\ 1 \end{pmatrix}, \\
\label{wavefunction2} \Psi_2=&\frac{1}{\sqrt{2}}(C_1^+ e^{ik_1 x}+
C_{1}^- e^{-ik_1
x}) \vec{\xi}_1+ \\
\nonumber
&+\frac{1}{\sqrt{2}}(C_{2}^+ e^{ik_2 x}+ C_2^- e^{-ik_2 x}) \vec{\xi}_2, \\
\label{wavefunction3} \Psi_3=&te^{ik x} \frac{1}{\sqrt{2}}\begin{pmatrix} 1 \\ 1 \end{pmatrix} + D e^{-\gamma x}\frac{1}{\sqrt{2}}\begin{pmatrix} 1 \\ -1 \end{pmatrix},
\end{eqnarray}
where $r$ and $t$ are the reflection and transmission amplitudes,
$k=\sqrt{2mE/\hbar^2}$,
$\gamma=\sqrt{2m\left(\Delta_{LT}-E\right)/\hbar^2}$ with $E$ being
the energy of the polaritons counted from the bottom of the lowest
TM mode in the waveguides. $C_{1,2}^{+(-)}$ are the complex amplitudes of
forward (backward) running waves in the active region with different
polarisations and wavevectors $k_{1,2}=\sqrt{2m(-U_0\pm\Omega/2+E)/\hbar^2}$. The vectors $\vec{\xi}_{1,2}$ read for scheme 1

\begin{eqnarray}
\vec{\xi}_1=\begin{pmatrix} 1 \\ 0 \end{pmatrix}, \vec{\xi}_2=\begin{pmatrix} 0 \\ 1 \end{pmatrix}
\end{eqnarray}
and 

\begin{eqnarray}
\vec{\xi}_1=\frac{1}{\sqrt{2}}\begin{pmatrix} 1 \\ i \end{pmatrix}, \vec{\xi}_2=\frac{1}{\sqrt{2}}\begin{pmatrix} 1 \\ -i \end{pmatrix}
\end{eqnarray}
for scheme 2. The reflection and transmission amplitudes can be then
found using the standard conditions of continuity of the wavefunctions and the current at the interfaces between the regions 1 and 2 and 2 and 3.

\begin{figure}
\begin{center}
\includegraphics[width=0.99\linewidth]{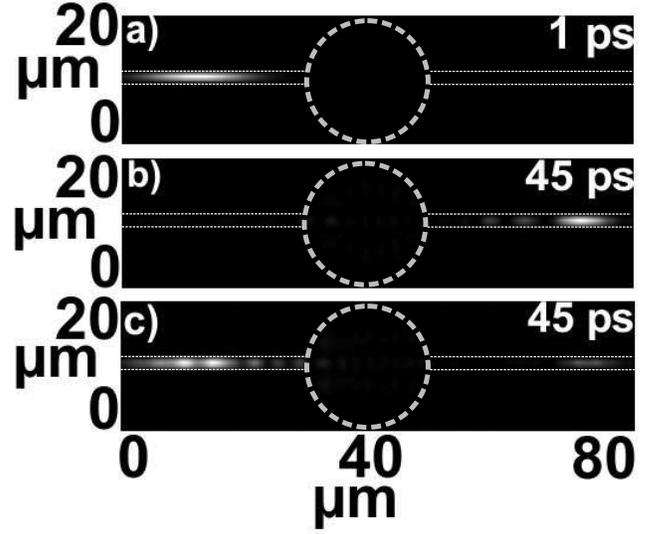}
\caption {\label{GP} Calculated real-space images of the intensity of the photon emission. Dashed lines indicate the potential profile. (a) The initial polariton pulse created by an external laser; (b) Transistor open ($B_{eff}=0$), the pulse passes through; (c) Transistor closed ($B_{eff}=0.2$ meV), the pulse is reflected. }
\end{center}
\end{figure}

Fig.\ref{fig1} (b) shows the dependence of the transmission coefficient
$T=|t|^2$ and reflection coefficient $R=|r|^2$ on the density of the optically pumped condensate for scheme 1 and Fig.\ref{fig1} (c) on $\Omega \propto V_g$ for scheme 2. The calculation is performed taking into
account realistic parameters of a GaAs microcavity. One sees that in both cases there is an
oscillating dependence of the outgoing polariton current on the
control parameter, and thus the system
pefectly matches the definition of a polariton spin transistor.

We have also performed a realistic numerical simulation of the
device operation. We describe the propagation of a linear polarized
pulse through the system with the use of coupled 2D spinor
Gross-Pitaevskii equation for excitons and Schroedinger equation for
photons \cite{Shelykh2006}, which for the case of the electrically controlled transistor (scheme 2) read

\begin{eqnarray}
i\hbar\frac{\partial\psi_\sigma}{\partial t}=\left[-\frac{i\hbar}{2\tau_{ex}}-\frac{\hbar^2}{2m_{ex}}\nabla^2+\alpha_1|\psi_\sigma|^2+\alpha_2|\psi_{-\sigma}|^2\right]\psi_\sigma-\\
\nonumber-i\sigma\Omega(\textbf{r})\psi_{-\sigma}+\frac{V_R}{2}\chi_\sigma,\\
i\hbar\frac{\partial\chi_\sigma}{\partial t}=\left[-\frac{i\hbar}{2\tau_{ph}}-\frac{\hbar^2}{2m_{ph}}\nabla^2+U_0(\textbf{r})\right]\chi_\sigma+\\
\nonumber+\sigma\Delta_{LT}(\textbf{r})\chi_{-\sigma}+\frac{V_R}{2}\psi_\sigma
\end{eqnarray}
where $\textbf{r}=\textbf{e}_xx+\textbf{e}_yy$, $\psi_\sigma,\chi_\sigma$ with $\sigma=\pm$ correspond to the excitonic and photonic fields written in basis of circular polarizations, $m_{ex}$ and $m_{ph}$ are effective masses of 2D excitons and photons respectively, $\tau_{ex}$ and $\tau_{ph}$ are their lifetimes, $V_R$ is the Rabi splitting, $\alpha_{1,2}$ are two constants characterizing polariton- polariton interactions, $U_0(\textbf{r}),\Delta_{LT}(\textbf{r})$ and $\Omega(\textbf{r})$ are profiles of the in- plane confinement potential for the photons, splitting  between the linear polarizations of the photonic mode and effective magnetic field in the active region respectively. 
 The waveguides have a thickness of $1
\mu$m, and the radius of the active region is $16 \mu$m. In this
region the pulse is rotated by an effective magnetic field induced
by the voltage applied to the gate electrode. The transition between
the waveguides and the active region has to be sufficiently smooth
in order to minimize the diffraction effects, which would decrease
the transmission coefficient in the $\Omega=0$ case. In our calculations we have considered photon lifetime $\tau_{ph}=$16 ps, exciton lifetime $\tau_{ex}$ 100 ps, LT splitting in the leads $\Delta_{LT}=$4 meV, Rabi splitting $V_R$=8 meV.

Figure \ref{GP} shows the results of the numerical simulations - the calculated real-space images of the photon emission. Panel (a) shows the gaussian pulse ($k=0.7\times 10^6$ m$^{-1}$) created by an external laser. The pulse should be sufficiently large in real space, in order to avoid the dispersion effects. The pulse propagates through the system quite rapidly, and after 45 ps leaves the active region through the waveguide. Panel (b) shows the $\Omega=0$ case (100\% transmission) and panel (c) corresponds to $\Omega=0.1$ meV, which reduces the transmission to approximately 20\%. The potential profile is indicated by the dashed lines.

Due to the rapid propagation of the pulse in the considered device, its operation time will be limited by the time needed to tune the values of the effective magnetic field in the active region. For the electrically controlled transistor this time is determined by the characteristic time of the tuning of the gate voltage $V_g$ and thus should be of the same order of magnitude as characteristic operation time for a standard Datta and Das transistor for the electrons. For optically controlled transistor (scheme 1), this time is determined by the time of the formation of stationary polariton distribution under the effect of the coherent circular polarized pump. The latter can be simply estimated by the longest characteristic time in the system, which is the exciton lifetime of the order of 100 ps in GaAs structures. This corresponds to an operation frequency of 10 GHz. Note also, that the recent proposal of engineering of the polariton confinement using a deposition of the metallic mesas of needed geometry \cite{Kaliteevskii2009} allows creation of polariton integrated circuits consisting from several transistors of the considered geometry.

In conclusion, we proposed two novel schemes of a polaritonic
analogue of Datta and Das spin transistor, which does not need
magnetic fields for operation. The proposed geometry allows to solve
the problems of decoherence and inefficient spin injection which
were blocking the experimental implementation of Datta and Das spin
transistor for electrons. The device transmissivity is easily and
quickly controlled tuning the intensity of the pumping laser or simply the gate
voltage. 

I.A.S. acknowledges the support from RANNIS "Center of excellence in polaritonics" and FP7 IRSES project "SPINMET".

\end{document}